\documentclass[12pt]{article}
\newcommand{\beq}{\begin{equation}}
\newcommand{\eeq}{\end{equation}}
\oddsidemargin
\topmargin
\usepackage{graphicx}
\usepackage{graphics}
\usepackage{latexsym}
\begin{document}
\vspace{1cm}
\begin{center}
{\Large \bf  Spatiotemporal intermittency and scaling laws in coupled map lattices}\\
\vspace{1cm}
Neelima Gupte$^{1}$ and Zahera Jabeen$^{2}$\\

Department of Physics, IIT Madras, Chennai 600036, India.\\
\end{center}

\begin{abstract}

We discuss the spatiotemporal intermittency (STI) seen in coupled map lattices
(CML-s). We
identify the types of intermittency seen in such systems in the context
of several specific CML-s. The Chat\'e-Manneville CML is introduced and
the on-going debate on the connection of the spatiotemporal
intermittency seen in this model with the problem of directed
percolation is summarised. We also discuss the STI seen in the sine
circle map model and its connection with the directed percolation
problem, as well as the inhomogenous logistic map lattice which
shows the novel phenomenon of spatial intermittency and other
types of behaviour not seen in the other models. The connection of the
bifurcation behaviour in this model with STI is touched upon. We
conclude with a discussion of open problems.

\end{abstract}

\section{Introduction}

Regimes of spatiotemporal intermittency,  wherein laminar regions, which exhibit regular dynamics in space and time, coexist and propagate together with  regions which show irregular or chaotic bursts, can be seen in a wide variety of spatially extended systems in the laboratory as well as in theoretical model systems. Such regimes have been regarded as precursors to regimes of fully developed spatiotemporal chaos \cite{Kaneko}. Spatiotemporal intermittent behaviour has been seen in theoretical models such as coupled map lattices \cite{kaneko1,Bohr},
probabilistic cellular automata \cite{Domany}, and  partial differential equations \cite{GL,St},
 as well as in
experimental systems such as chemical reactions \cite{Kapral}, Rayleigh-Benard convection in narrow channels and annuli \cite{RB,RB1}, planar Couette flow \cite{PC}, studies of fluid flows between rotating eccentric cylinders such as the Taylor-Dean \cite{TD} and Taylor-Couette \cite{Colovas,CT} flows, and the `printer's instability' \cite{printer}.
A variety of scaling laws have been observed in these systems. However, there
are no definite conclusions about  their
universal behaviour. Many of the observed phenomena have been seen in experimental systems where no simple model is available. 
There is a lively on-going debate about the nature of spatiotemporal intermittency and its analogy with systems which undergo phase transitions.
Thus, the analysis of spatiotemporal intermittency remains a challenging theoretical problem. 
In this review, we will concentrate on spatiotemporal intermittency in coupled map lattices, i.e. systems with continuous variables which evolve on discrete space-time.

Spatiotemporally intermittent behavior has been divided into two classes.
In the first type, there is no spontaneous creation of bursts. 
Under spatial coarse-graining, if a site and its neighbours show laminar behaviour at a given step, the site remains laminar 
at the next time step. However it can show turbulent behaviour if infected by a turbulent neighbour. A turbulent site can either relax to laminar behaviour or infect its neighbour. Below a threshold parameter, which controls the propagation rate of turbulence, the turbulence dies out. Once the whole system reaches the laminar state, it has no way of escaping from this state. Hence the laminar state is called the absorbing state. Above the threshold parameter, the turbulence spreads over the whole lattice. Thus the burst states `percolate' through space-time. It has therefore been argued that the transition to this type of 
spatiotemporal intermittency is a 
second order phase transition, and that this transition falls in the same universality class as directed percolation \cite{Pomeau} with the laminar states being identified with the `inactive' states and the turbulent states being identified as the `active' or percolating states. This conjecture has become the central issue in a long-standing debate 
\cite{Rolf,Chate,Houlrik,Grassberger,Janaki}, which is still not completely resolved.

In the second type of intermittency, the spontaneous creation of bursts can be seen, given some coarse-grained reduction of the states. Thus, for this case, a laminar site has a finite probability of undergoing  a turbulent burst even if all its neighbours are laminar. 
This kind of intermittency is associated with the transition of a
spatial pattern. A typical example of this type of intermittency is
observed in the coupled logistic lattice from pattern selection to fully
developed spatiotemporal chaos \cite{kaneko1} as well as in experiments
on Benard convection \cite{RB,RB1}, Faraday instabilities \cite{Faraday}, 
and in two-dimensional electric convection in a liquid crystal 
\cite{electroconvection}.

\section{The Chat\'e-Manneville map and its variants:}

The conjecture by Pomeau that the transition to spatiotemporal intermittency of the first type with a laminar absorbing state belongs to the same universality class
as directed percolation has been very extensively investigated. A very simple coupled map model was introduced by Chat\'e and Manneville for the investigation of this conjecture.

The Chat\'e-Manneville maps at different sites are coupled together by linear
diffusive coupling of the form
\beq
x^{n+1}(i)= (1-\epsilon)f(x^{n}(i)) + \frac{\epsilon}{2}[f(x^n(i+1))+f(x^n(i-1)) ]
\eeq

The single site Chat\'e-Manneville map has the form
\begin{eqnarray}
f(x) & = & r x  \hspace{0.5in} 0 \leq x < 1/2  \nonumber\\ 
     & = & r(1-x) \hspace{0.5in} 1/2 \leq x \leq 1 \nonumber \\
     & = & x      \hspace{0.5in}  1 < x \leq r/2 \nonumber\\
\label{ChatCML}
\end{eqnarray}
where $r > 2$. The return map is shown in Fig. 1. 

The uncoupled dynamics of such a map is chaotic as long as $f(x)$ remains in the unit interval since $r > 1$ everywhere in this domain. However, as soon as $f(x) > 1$ the iteration is locked on to a fixed point. The local phase space is thus a marginally stable fixed line  ( $x \in [1, r/2]$), connected to a chaotic repeller ( $x \in [0,1]$). There is thus a clear and natural distinction between laminar and turbulent states for a given site. Chat\'e and Manneville argued that since the model has no single stable fixed point or globally attracting period, there is no additional local dynamics to complicate the transition from local transient chaos to global spatiotemporal chaos via the spatial coupling alone.
    
The dynamics of this chain was studied with random initial conditions. At $\epsilon > \epsilon_{c}$, sustained spatiotemporal intermittency was seen for the system for $r=2.1$ and $r=3$. The lattice showed regions of laminar behaviour punctuated by chaotic bursts (See Fig. 2).
The distribution of the  lengths of the laminar regions showed power-law behaviour of the
form  $P(l) \approx l^{-\zeta}$ with the exponent $\zeta$ taking values between $1.67$ and $1.78$ at $r=2.1$ and $1.9$ to $1.99$
at $r=3$.(The simulated DP values for the corresponding exponent ranged from $1.6$ to $1.74$). It was initially argued that the values obtained showed significant departures from the directed percolation exponents and hence that spatiotemporal intermittency did not belong to the DP universality class. However Grassberger et al examined the same CML later \cite{Grassberger} and found that the departures from DP exponents at $r=2.1$ were insignificant on long time-scales. In contrast, statistically significant departures from DP were observed at $r=3.0$.
It was observed that for this case, long straight lines of turbulent sites were seen in the space-time plot (Fig. 2(b)). These were called `solitons' and were
reflective of long range correlations in the system. Grassberger and Schreiber argued that these long range correlations spoilt the analogy with DP and hence 
led to the departure of the STI in this system from the DP universality class.
The existence of solitons led to the emergence of a length and time scale (the mean life time of the solitons) which is independent of critical fluctuations, and hence led to departures from the DP class \cite{Grassberger}.
Since the solitons in this model had very long life-times, they also conjectured that the cross-over times in this model could be very long and
the real asymptotic behaviour of the model could still be in the DP universality class.

The conjecture that the solitons were the primary ingredient which
spoilt DP universality was tested in a variety of ways \cite{Rolf,
Bohr1, Mikkelson}. 
The CML of eq. ~\ref{ChatCML} was evolved with asynchronous up-dates i.e. the maps at every site were not updated simultaneously, but were updated one after another in a random sequence. The asynchronous updates served to destroy the solitons and good agreement with a complete set of DP exponents was obtained (see Appendix for definitions of the DP exponents and Table 1 for obtained values). 

Another study introduced a deterministic extension  of the
Chat\'e-Manneville model, where the soliton properties could be tuned \cite{Bohr1, Mikkelson}.  
It was seen that the role played by the solitons in the transition to turbulence was more profound than that of setting a cross-over time. They were able to change nature of the transition from second to first order. 

The CML equations of this solitonic Chat\'e-Manneville model were
\begin{eqnarray}
x^{n+1}(i) & = & (1-\epsilon)f(x^{n}(i)) + \frac{\epsilon}{2}[f(x^n(i+1))+f(x^n(i-1)) ] + y^{n}(i)  \nonumber \\
y^{n+1}(i) & = & b(x^{n+1}(i) - x^{n}(i)) \nonumber \\
\label{CML2}
\end{eqnarray}
where the function $f(x)$ is the same as that in Eq.~\ref{ChatCML}. The map becomes 
increasingly two dimensional with increase in $ |b| $. Thus if $x^*$ is a fixed
point of the old CML Eq. ~\ref{ChatCML}, it gets mapped to the fixed point $(x^*,0)$ of Eq. ~\ref{CML2}. The CML thus has three adjustable parameters $r,\epsilon, b$. The 
parameter $r$ was set to $3.0$ and the phase diagram in $\epsilon, b$ space was obtained. The parameter $b$ has a large influence on the existence of solitons.
The $b=0$ case is the original Chat\'e-Manneville case where there are long lived solitons (Fig. 3 (c),(d)). These life-times become even longer when $b=-0.1$ (Fig. 3(a),(b)), and the solitons only vanish when they collide with other solitons or propagate into turbulent structures (Fig. 3(e),(f)). The order parameter $m$ which is the number of active or turbulent sites, jumps sharply as a function of $\epsilon -\epsilon_c$ as in a first order transition \cite{Bohr1, Mikkelson}. The behaviour of $m$ at  $b=0.0$ and $b=0.2$ is consistent with a continuous  transition with exponents that approach DP values more closely as 
$|b|$ increases corresponding to soliton-poor regimes (Table. 2). Thus not only does the existence of solitons spoil the analogy of spatiotemporal intermittency with DP, but it can change the order of the transition itself.
Thus, in order to obtain  a clean comparison between directed percolation
and spatiotemporal intermittency, it is desirable to examine a deterministic 
system which is free of coherent propagating structures.
We describe such a
system in the next section.

\section{ The sine circle coupled map lattice:}

     The sine circle map lattice is defined by
\beq
\theta^{n+1}(i)= (1-\epsilon)f(\theta^{n}(i)) + \frac{\epsilon}{2}[f(\theta^n(i+1))+f(\theta^n(i-1)) ] ~~~(mod ~~1)
\eeq
with periodic boundary conditions. The local map is the sine circle map
\beq
f(\theta^n)=\theta^n + \Omega - \frac{K}{2\pi}\sin (2\pi \theta^n)
\eeq

Here, $K$ is the strength of nonlinearity, $\Omega$ is the frequency of each single sine circle map and, $\epsilon$ is the coupling strength. This system is a popular model for the behaviour of mode-locked oscillators. The system shows regimes of spatiotemporal intermittency
at certain points in the parameter space when random initial conditions are evolved synchronously. The phase diagram of the system at $K=1$ is shown in Fig. 4  and the 
points where spatiotemporal intermittency are seen are marked with diamonds \cite{Janaki} and pluses  \cite{zahera}.
This system shows a transition to spatiotemporal intermittency in the
presence of two types of absorbing states.
The two qualitatively distinct
absorbing regions are as follows \cite{Janaki}.

(i) When the nonlinearity parameter $K = 1$, there are regions of
$(\epsilon - \Omega)$ space where the system goes to the synchronised
spatiotemporal fixed point $\theta^{\star} = \frac{1}{2\pi}
\sin^{-1}(\frac{2\pi \Omega}{k})$. This constitutes an  unique
absorbing state. The critical
behaviour was examined at 2 critical points in this regime: $\Omega=0.064,
\epsilon=0.63775$, and $\Omega=0.068, \epsilon=0.73277$. These mark
the transition from a laminar phase to STI. The turbulent sites here
are those which are different from $\theta^{\star}$.

(ii) When the nonlinearity parameter $K = 3.1$, there are regions of
$(\epsilon - \Omega)$ space where sites with any value less than $1/2$
constitute the absorbing states, and sites whose values are greater
than $1/2$ constitute the turbulent states. 
The absorbing states are  now infinitely many, as also  weakly
 chaotic.  The critical behaviour was examined at 2 critical points in
this regime as well: $\Omega=0.18, \epsilon=0.701$, and $\Omega=0.19,
\epsilon=0.65612$.

The exponents for the onset of spatiotemporal intermittency have been defined both for the bulk critical exponents and the spreading exponents (See Tables 3 and 4). In both cases there is good agreement with DP values \cite{FN}. Thus 
the coupled sine circle map constitutes a deterministic system evolving 
under synchronous up-dates where DP exponents are straight-forwardly obtained.
There are no long-lived coherent structures at any of the parameter values where the system has been studied to spoil the analogy with DP. This also lends credence to the argument that it is the solitons which lead to the departure of 
the Chat\'e-Manneville exponents from the DP class.
Certain parameter regions of the sine circle map model show regimes of
spatiotemporal intermittency which is unlike the DP behaviour. These 
are currently under investigation. Such regimes have been seen earlier
in 
a coupled map model which we will discuss below.
 
\section{The inhomogeneous logistic map lattice:}

The specific system studied by us consists of an inhomogeneous
lattice of coupled maps where the inhomogeneity appears in the form of
different values of map parameter at different  sites.
Such lattices have been considered
in the case of pinning studies \cite{pinn}.
The coupling between maps is diffusive and nearest
neighbour.  We consider the simplest case, that is, a situation where neighbouring lattice sites have different values of the parameter and alternate sites
have the same value of the parameter.

This  model is defined by the evolution equations 
\begin{equation}
 x^{n+1}(2i)=(1-\epsilon)f(x^{n}(2i), \mu)+{\epsilon\over
2}{[f(x^{n}(2i-1),
     \mu^{\prime}) +f(x^{n}(2i+1), \mu^\prime)]} 
\label{cml} 
\end{equation}

where $f(x)= \mu x(1-x)$ is the logistic map and $\mu$, $ \mu^\prime \in$
 $[0,4]$, $ x^{n}(2i)$ is the value of the variable $x$ at the even lattice site
$2i$ at time $n$, and $ 0 \leq x \leq 1$. In the case of $x^{n}(2i+1)$, the variable defined at odd lattice sites, the evolution equation is defined by the evolution above with $2i$ being replaced by $2i+1$ and $\mu$ and $\mu^ \prime$ interchanged.
We set $\mu^\prime=\mu-\gamma$ and  use periodic
                   boundary conditions where $2N$, the number of lattice sites is even.
The synchronised fixed points of the system are given by $x^{*}=0$ and
$x^{*}={{\mu-\gamma\epsilon-1}\over{\mu-\gamma\epsilon}}$.

Spatiotemporal intermittency can be found at many points in the
parameter space, unlike the homogeneous logistic coupled map lattice, where STI is found for values of
$\mu$ which lie near the period $3$ window.

Bifurcations  from the synchronised fixed points can be used to obtain indicators of regions 
where spatiotemporal intermittency can be found. 
Spatiotemporal intermittency can be seen in the vicinity of bifurcation
lines of co-dimension $1$ and bifurcation points of co-dimension $2$
where such lines intersect each other.
It is interesting to
note that pure spatial intermittency, where there are  bursts
interspersed with laminar behaviour on the spatial
lattice, accompanied by purely temporal periodic behaviour, can be found in the vicinity of some
tangent-period doubling points where the eigen-values of the stability matrix
cross $+1$ and $-1$ at the same point in the parameter space. 

These  conditions are satisfied for the fixed point $x^{*}=0.0$ in the neighbourhood of
$\epsilon=0.635$, $\gamma=1.1$
where the eigen-values are
$ -0.9127$, and   $1.0585$ respectively.
We show the spatial intermittency present in the vicinity of this point in Fig. 5. The parameter values chosen are $\gamma=1.19$ and $\epsilon=0.635$.
 The space-time plot of the system can be seen in Fig. 5. The temporal period two structure of both the laminar and burst sites can be seen clearly in the space-time plot.
It is clear that the time evolution of the lattice shows stable
period two oscillations at both the laminar and the burst sites.  Thus
the intermittency is purely spatial in nature and
temporally we have stable periodic behaviour.

The temporally periodic spatial structure seen here shows the unusual occurrence of long range
spatial correlations.
This is reflected  in the distribution of laminar lengths.
The length of the laminar lengths, i.e. the
number of consecutive sites which follow periodic behaviour before being
interrupted by bursts is calculated. As mentioned above  the spatial period in the laminar regions  is two.
The distribution of laminar lengths shows power-law behaviour $P(l) \approx l^{-\zeta}$ with exponent $\zeta=1.1$.

Several other tangent-period-doubling points are also seen in this parameter space and spatial intermittency with an associated scaling exponent $\zeta=1.1$ can be seen in their vicinity , except for the tangent-period doubling point at $\gamma =0.66$ and $\epsilon=0.39$ where spatiotemporal intermittency
is seen (See Fig. 6(a),(b)). The distribution of laminar lengths seen here
also shows power-law scaling $P(l)=l^{-\zeta_F}$ but now with exponent $\zeta_F=2.0$. 
Spatiotemporal intermittency can also be seen along lines of co-dimension 1
bifurcations along the parameter space. Here, other exponents $\zeta_2=1.33$ and
$\zeta_3=1.66$ are seen. These appear to be crossover exponents. 

It is interesting to note that an exponent which
falls within the same range as $\zeta_F$ has been seen in the case of
Rayleigh-Benard convection in an annulus \cite{RB1}, 
exponents that fall in the same range as $\zeta_2$ have been seen for
convection in a channel and for the Taylor-Dean experiment, and
$\zeta_3$ has been seen in the Chat\'e-Manneville CML at $r=2.1$ and
$\epsilon=0.0045$ as well as in DP simulations \cite{Colovas}. Similarly the Chat\'e-Manneville CML at $r=3.0$
$\epsilon_c=0.360$
 shows the exponent $\zeta_F$.
The behaviour in this model is complicated by the fact
that the laminar state is periodic, and that a bifurcation to period two
seems to take place very close to the bifurcation from the synchronised
fixed point, therefore it is difficult to draw direct connections to  behaviour
seen in the earlier models. 
However, the model provides pointers to the following conjectures.
Spatiotemporal intermittency can be found in the vicinity of
bifurcations from low-ordered periodic points. The type of bifurcation
that takes place as well as the nature of the stable state which
precedes the bifurcation may provide an indication of the universality
class of the spatiotemporal intermittency. Further work on this 
system could lead to interesting results.

\section{Discussion}

To summarise, spatiotemporal intermittency is a dynamical regime with interesting 
but little understood behaviour. While a plethora of phenomenological
information is available on the behaviour of systems which show STI,
both theoretical and experimental,
a unifying picture is lacking.  While the conjecture that several
systems which demonstrate spatiotemporal intermittency belong to the
DP universality class finds some support, further work is necessary to 
identify the  factors which contribute to the presence or absence of DP.

The effect of travelling coherent structures on the transition to
spatiotemporal intermittency deserves to be explored further, as does 
the connection with bifurcation behaviour. Studies of systems where 
the laminar state is not absorbing are few and far between. Systematic  studies
of such systems need to be undertaken. 
Further work on experimental systems is also necessary. There has been
only one experiment where DP exponents have been found so far
\cite{Rupp}. Further work also needs to be done in constructing models
for the   
quasi-1d fluid systems which show spatiotemporal intermittency.
The extent to which the behaviour seen in stochastic models like
probabilistic cellular automata, is similar to the behaviour in strictly
deterministic models, like those considered here, also needs to be
explored further.
Thus, we expect this
area of research to remain active and contribute to many lively debates  in the coming years.

\newpage
\section{Appendix}

The quantities characterised by critical exponents at
$\epsilon=\epsilon_c$, the critical value at which the transition to 
sustained spatiotemporal intermittency takes place, are :
\begin{enumerate}

\item The static exponents:

 \begin{enumerate}
\item The escape time, $\tau$, i.e. the time taken by the system, starting from a random initial
state, to reach the absorbing state, shows power law behaviour of the form $\tau\sim L^z$.
\item The order parameter $m$, i.e. the fraction of turbulent sites, scales as $m(\epsilon_c,L,t)\sim(\epsilon-\epsilon_c)^{\beta}$, for
$\epsilon>\epsilon_c$, at fixed $t$.
\item For $t\ll\tau$,  $m(\epsilon_c,L,t)\sim t^{-\beta/{\nu z}}$, where
$\nu$ is the correlation length exponent in space.
\item The correlation function $C_j(t)$ is defined as
\begin{equation}
C_j(t)=\frac{1}{L}\sum_{i=1}^{L} < x_i(t)x_{i+j}(t) >-<x_i(t)>^{2} 
\end{equation}
where $x_i(t)$ is the variable value at site $i$ at time $t$. The brackets denote averaging over different initial conditions. This quantity scales as $C_j(t)\sim j^{1-\eta'}$.
\item These bulk exponents satisfy the hyperscaling relation
$2\beta/{\nu}=2-d+\eta'$.
\end{enumerate}

\item When the turbulent seeds spread in an absorbing lattice, the
quantities associated with the spreading exponents are  :

\begin{enumerate}
\item $N(t)$, the number of active sites at time $t$. This scales as $N(t)\sim t^{\eta}$,
\item $P(t)$, the survival probability, or the fraction of initial conditions which show a non-zero number of active sites at time $t$, which shows scaling behaviour of the form $P(t)\sim t^{-\delta}$.
\item  $R^2(t)$, the mean squared deviation of the position of the active sites from the original sites of turbulent activity, averaged over the surviving runs alone.This scales as $R^2 (t) \sim t^{z_s}$
\end{enumerate}

\end{enumerate}
\newpage
\begin{center}
\begin{center}
\begin{table}
\begin{center}
\begin{tabular}{|c|c|c|c|c|}
\hline
$r$ & $\epsilon_c$ &  $z$   &  $\beta$  & $\beta/ \nu z$ \\ \hline
2.2 & 0.0195(2)  & 1.58(2) & 0.28(1) & 0.16(1) \\
3.0 & 0.5870(3)  & 1.60(3) & 0.28(1) & 0.15(2) \\ \hline
DP  &            & 1.57    & 0.28    & 0.16   \\ \hline 
\end{tabular}
\vspace{0.5cm}
\caption{The critical exponents of the asynchronously updated Chat\'e-Manneville coupled map lattice. The last row shows the corresponding DP exponents. }

\end{center}
\vspace{1.0in}
\end{table}
\end{center}

\begin{center}
\begin{table}
\vspace{0.5cm}
\begin{center}
\begin{tabular}{|c|c|c|c|c|}
\hline
$r$ & $b$ & $\epsilon_c$ & $z$ & $\beta/\nu z$\\ \hline
3.0 & -0.1   & 0.35203(1) & 1.52(3) & 0.02(2) \\
    &  0.0   & 0.35984(3) & 1.42(2) & 0.18(1) \\
    &  0.2   & 0.37323(1) & 1.58(1) & 0.16(1)\\ \hline
DP  &        &            & 1.58074 & 0.15947\\ \hline 
\end{tabular}
\vspace{0.5cm}
\caption{The critical exponents of the extended 2 dimensional Chat\'e Manneville CML for different values of  $ b $. $b= 0.2$ shows exponents closer to the DP exponents.}
\end{center}
\end{table}
\end{center}
\end{center}
\begin{table}
\begin{center}
\begin{tabular}{|c|c|c|c|c|c|c|}
\hline
$k$ & $\Omega$ & $\epsilon$& $z$ &$\beta$ & $\nu$ & $\eta^{\prime}$\\
\hline
$1$ & $0.068$ & $ 0.63775$ & $ 1.580$ & $0.28$ & $1.10$ & $1.49$  \\
$1$ & $0.064$ & $0.73277$ & $1.591$ & $0.28$ & $1.10$ & $1.50$ \\
$3.1$ & $0.18$ & $0.70100$ & $1.597$ & $0.26$ & $1.12$ & $1.50$\\
$3.1$ & $0.19$ & $0.65612$ & $1.591$ & $0.28$ & $1.10$ & $1.49$\\
$ DP$ & $~~$ & $~~$ & $1.58$ & $0.28$ & $1.10$ & $1.51$\\
\hline
\end{tabular}
\end{center}
\caption[]{ Critical static exponents of the synchronously updated
coupled circle map lattice for 4 critical points. The first two critical
points correspond to transitions to an unique absorbing case, while
the third and fourth points correspond to weakly chaotic absorbing
states.
The last row shows the corresponding exponents of directed percolation.}
\end{table}

\begin{table}
\begin{center}
\begin{tabular}{|c|c|c|c|c|c|}
\hline
$k$ & $\Omega$ & $\epsilon$& $\eta$ &$\delta$ & $z_s$ \\
\hline
$1$ & $0.068$ & $ 0.63775$ & $0.292 $ & $0.153$ & $1.243$ \\
$1$ & $0.064$ & $0.73277$ & $0.302$ & $0.158$ & $1.259$\\
$3.1$ & $0.18$ & $0.70100$ & $0.310$ & $0.157$ & $1.272$ \\
$3.1$ & $0.19$ & $0.65612$ & $0.308$ & $0.156$ & $1.251$ \\
$ DP$ & $~~$ & $~~$ & $0.313$ & $0.159$ & $1.26$ \\
\hline
\end{tabular}
\end{center}
\caption[]{ Spreading exponents of the synchronously updated
coupled circle map lattice for 4 critical points. Two active seeds in
an absorbing configuration is used as initial condition. For the first
2 critical points there exists an unique absorbing state, while for
the third and fourth points one can have many different absorbing
states and consequently many different initial absorbing backgrounds.
However we notice that the exponents obtained are the same for
different initial preparations and thus appear universal for 2 active
seeds.  The last row shows the corresponding exponents of directed
percolation.}
\end{table}
\newpage

\begin{center}
\begin{figure}
\centering
\includegraphics[scale=1.0]{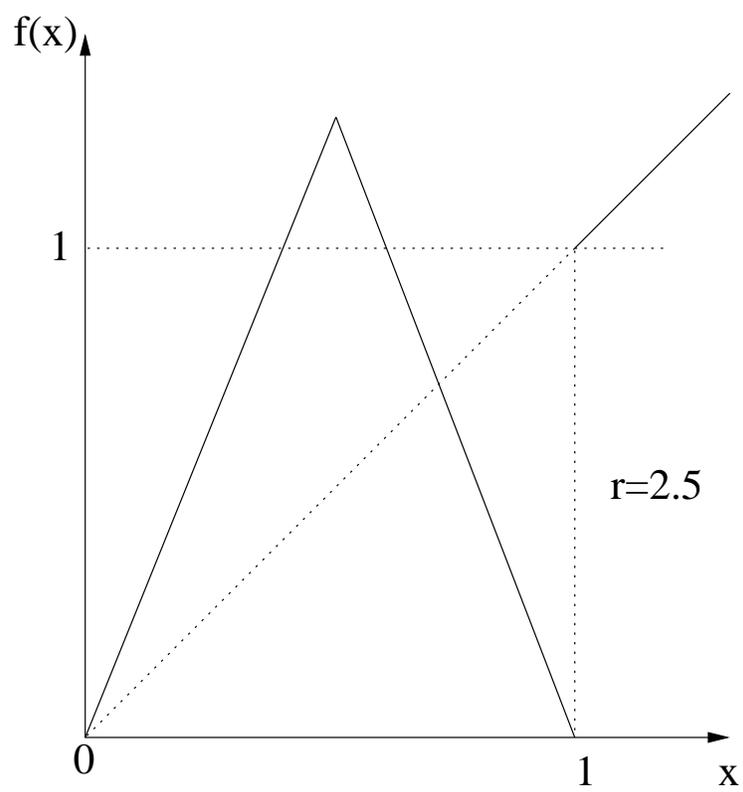}
\caption{The Chat\'e-Manneville map at $r=2.5$}
\end{figure}
\end{center}

\begin{center}

\begin{figure}
\begin{center}
\begin{tabular}{lc}
(a) &\includegraphics{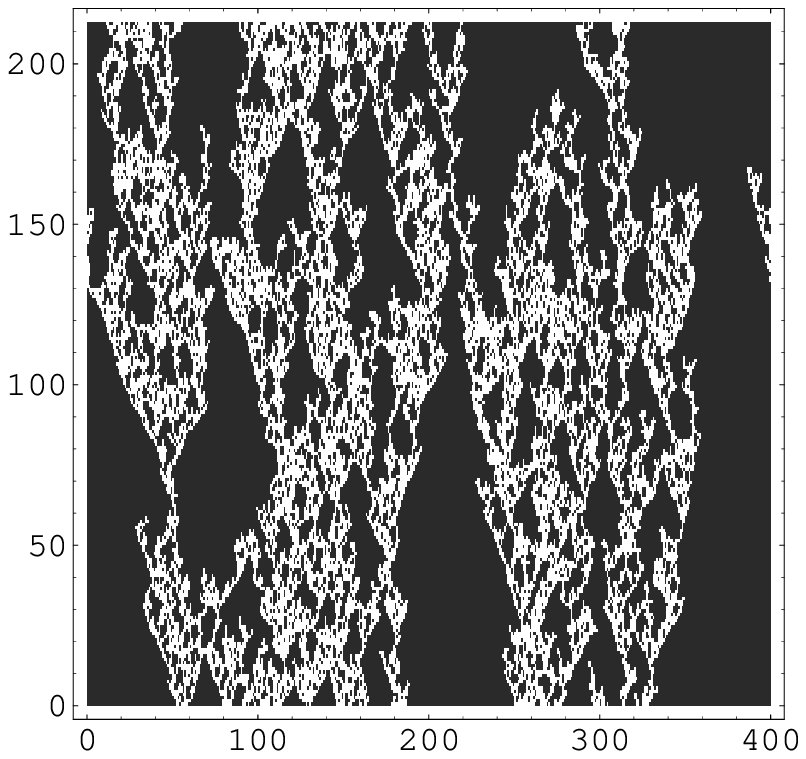}\\
(b) &\includegraphics{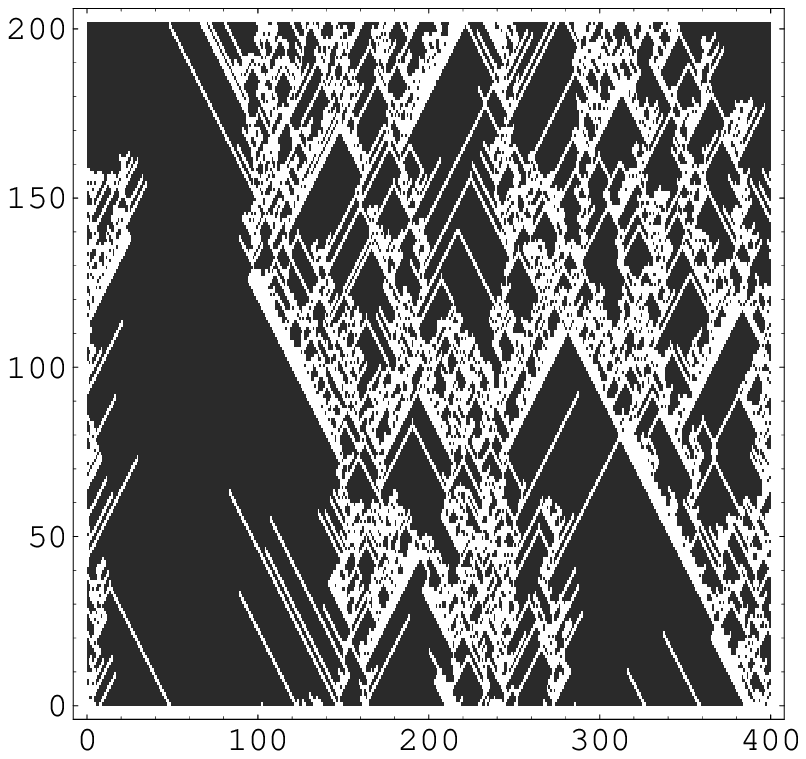}\\
\end{tabular}
\caption{The spacetime plots for the Chat\'e-Manneville map for (a) $r=2.1$, $\epsilon_c = 0.0045$ and (b)$ r = 3.0$, $\epsilon_c=0.36$ after discarding $3200$ transients. Laminar sites are the black spaces. In (a), every 8th timestep has been plotted. The slower dynamics at $r=2.1$ can be seen from the plot. Also, the solitons are clearly seen in (b).}
\end{center}
\end{figure}
\end{center}

\begin{center}
\begin{figure}
\begin{tabular}{cc}
 (a) \includegraphics[scale=0.8]{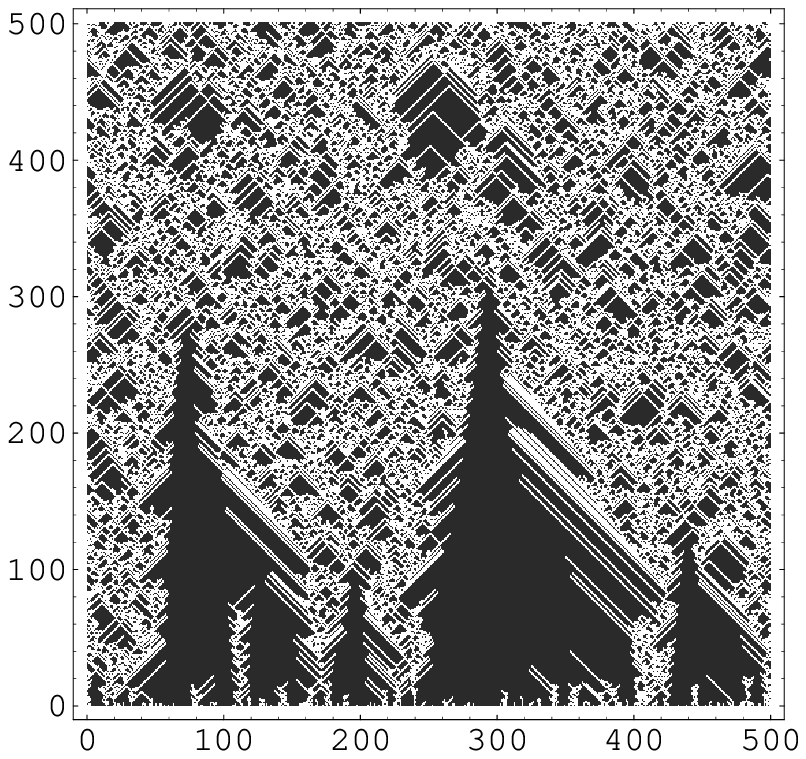}& (b) \includegraphics[scale=0.8]{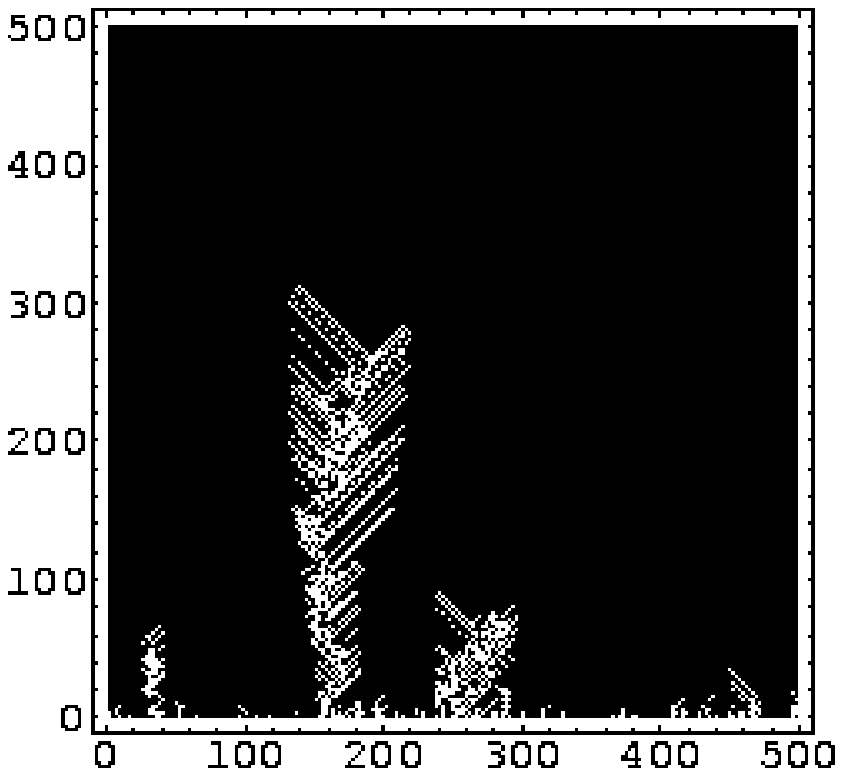}\\
(c) \includegraphics[scale=0.8]{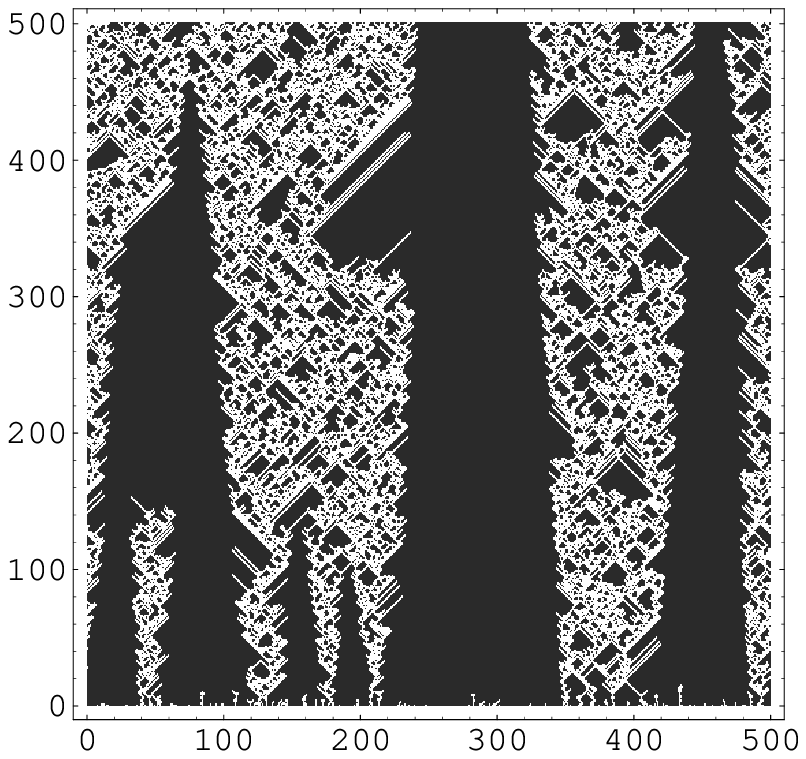}&(d)\includegraphics[scale=0.8]{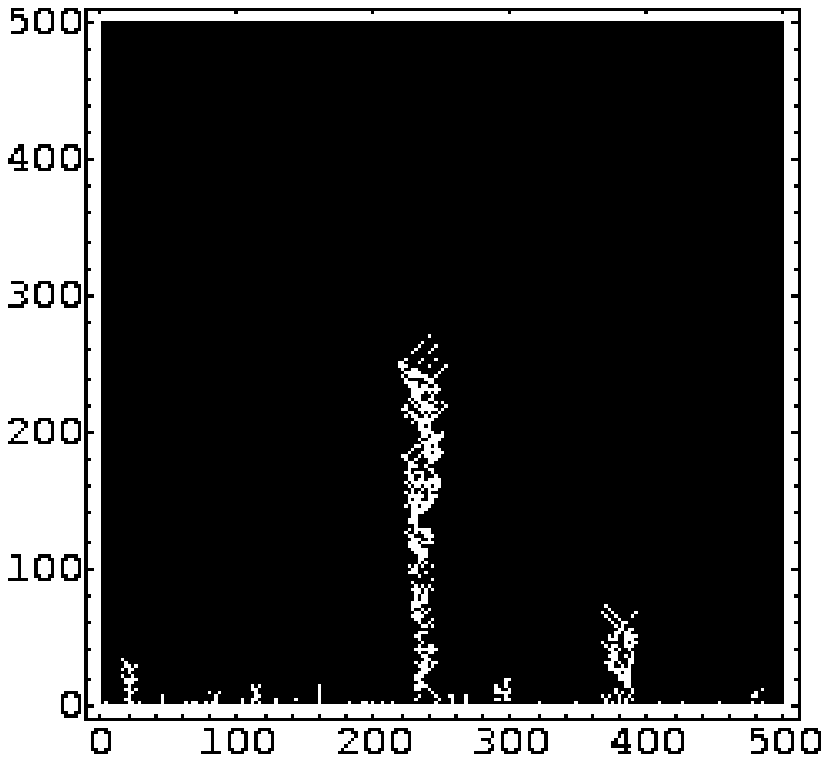}\\
(e)\includegraphics[scale=0.8]{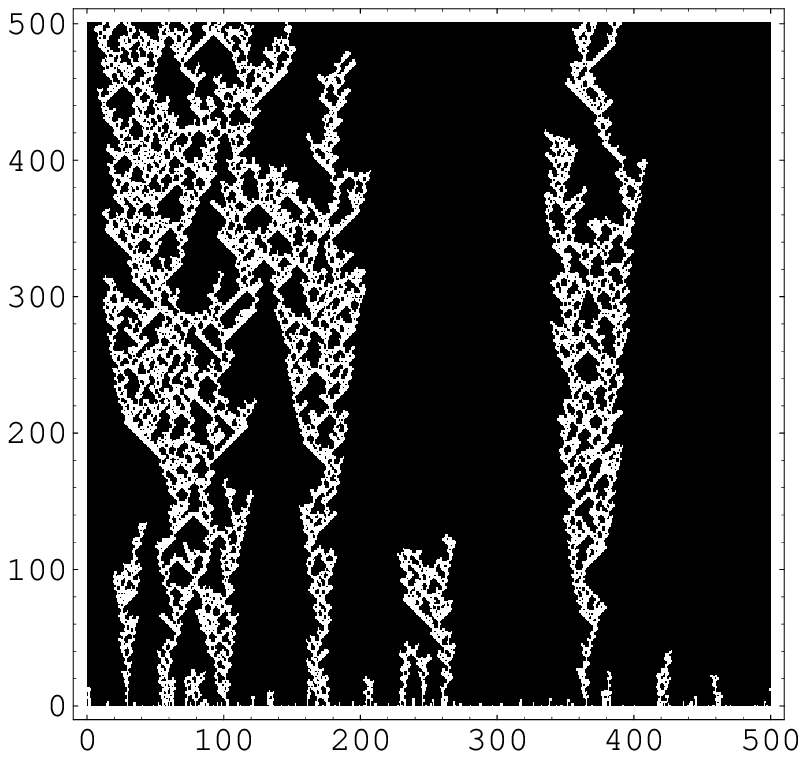}&(f)\includegraphics[scale=0.8]{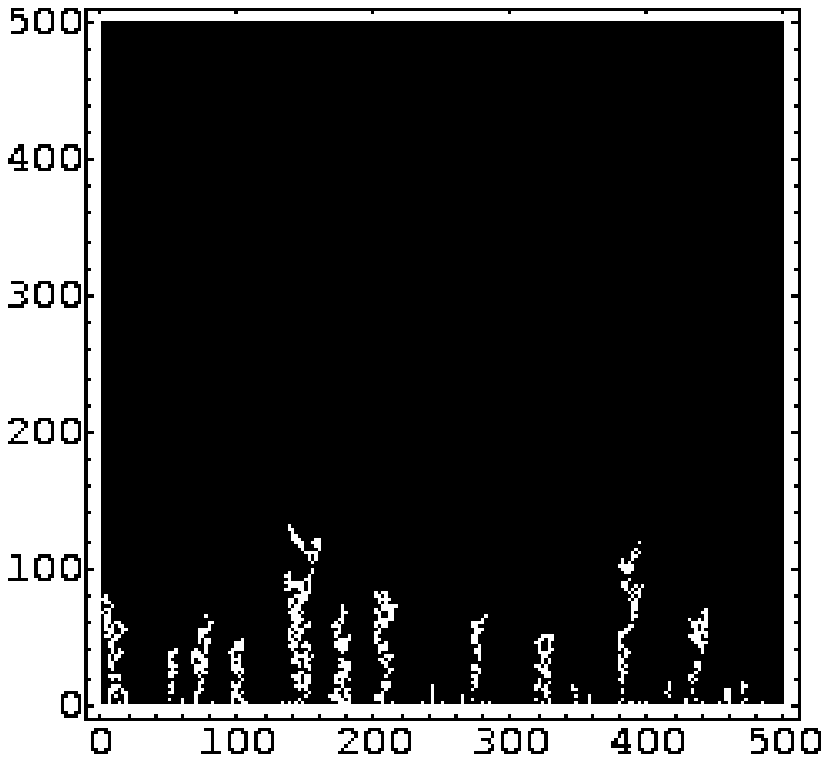}\\
\end{tabular}
\caption{Space-time plots at points above (left column) and below (right column) criticality for the 2 dimensional Chat\'e-Manneville map for different values of $ b $.  (a) $b=-0.1$, $\epsilon=0.359$ , (b) $b=-0.1$, $\epsilon=0.3537$, (c) $ b=0.0$, $\epsilon=0.370$ , (d) $b=0.0$, $\epsilon=0.365$ , (e) $b=0.2$, $\epsilon=0.392$ , (f) $b=0.2$, $\epsilon=0.384$ . Solitons are seen at $b=-0.1$ and $b=0.0$ whereas they are absent at $b=0.2$.   }
\end{figure}
\end{center}

\begin{figure}
\centering
\includegraphics[height=10cm,width=12cm]{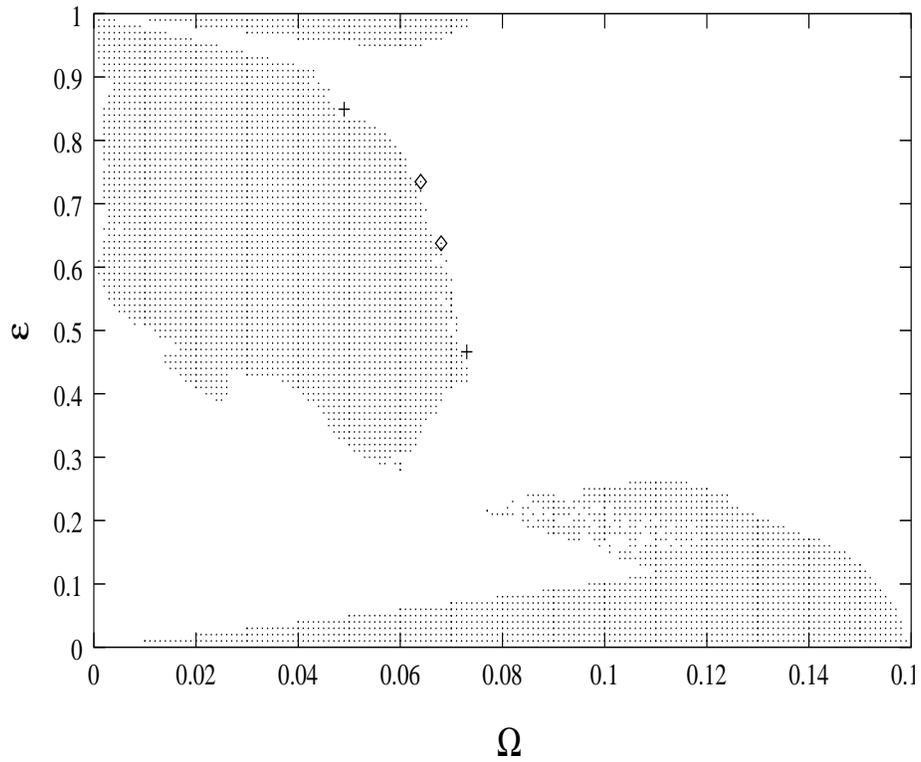}
\caption{The phase diagram in $(\epsilon,\Omega)$ space for the coupled circle map lattice at $K=1$ after discarding $15000$ transients for a lattice of $1000$ sites. Random initial conditions were used. The regimes where spatially synchronised and temporally fixed solutions are obtained are shown by dots. The points at which DP-like  exponents are obtained are marked by $\Diamond$ [20] and + [25].}
\end{figure}

\begin{center}
\begin{figure}
\centering
\includegraphics{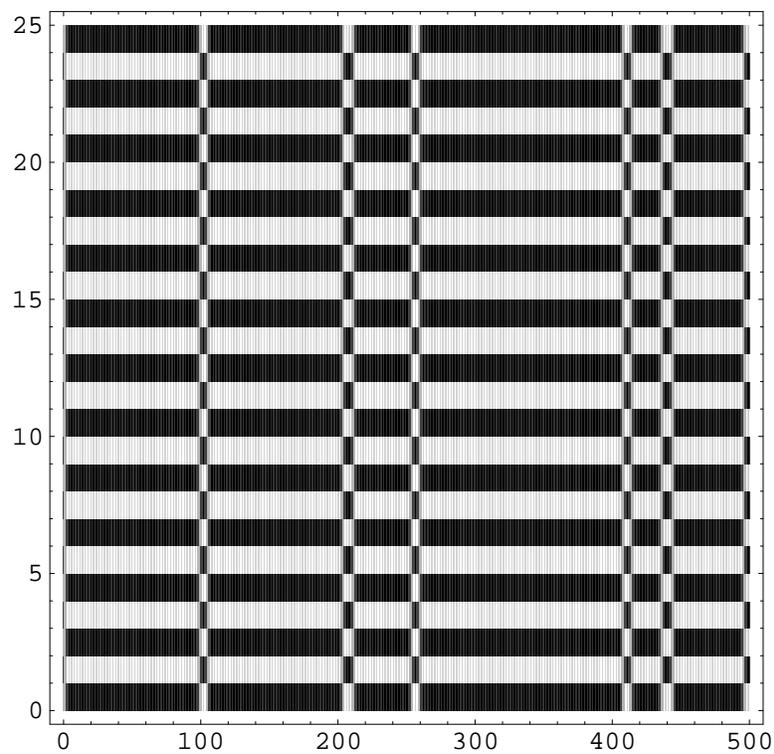}
\caption{Spacetime plot of the inhomogenously coupled logistic map at $\gamma=1
.19$ and $\epsilon=0.635$ for a system of 500 sites. The last 25 iterations are shown after discarding $5000$ transients. Pure spatial intermittency, where the laminar region is temporally periodic is seen.}
\end{figure}
\end{center}

\begin{center}
\begin{figure}
\begin{tabular}{cc}
(a)& \includegraphics[height=3.8in,width=4in]{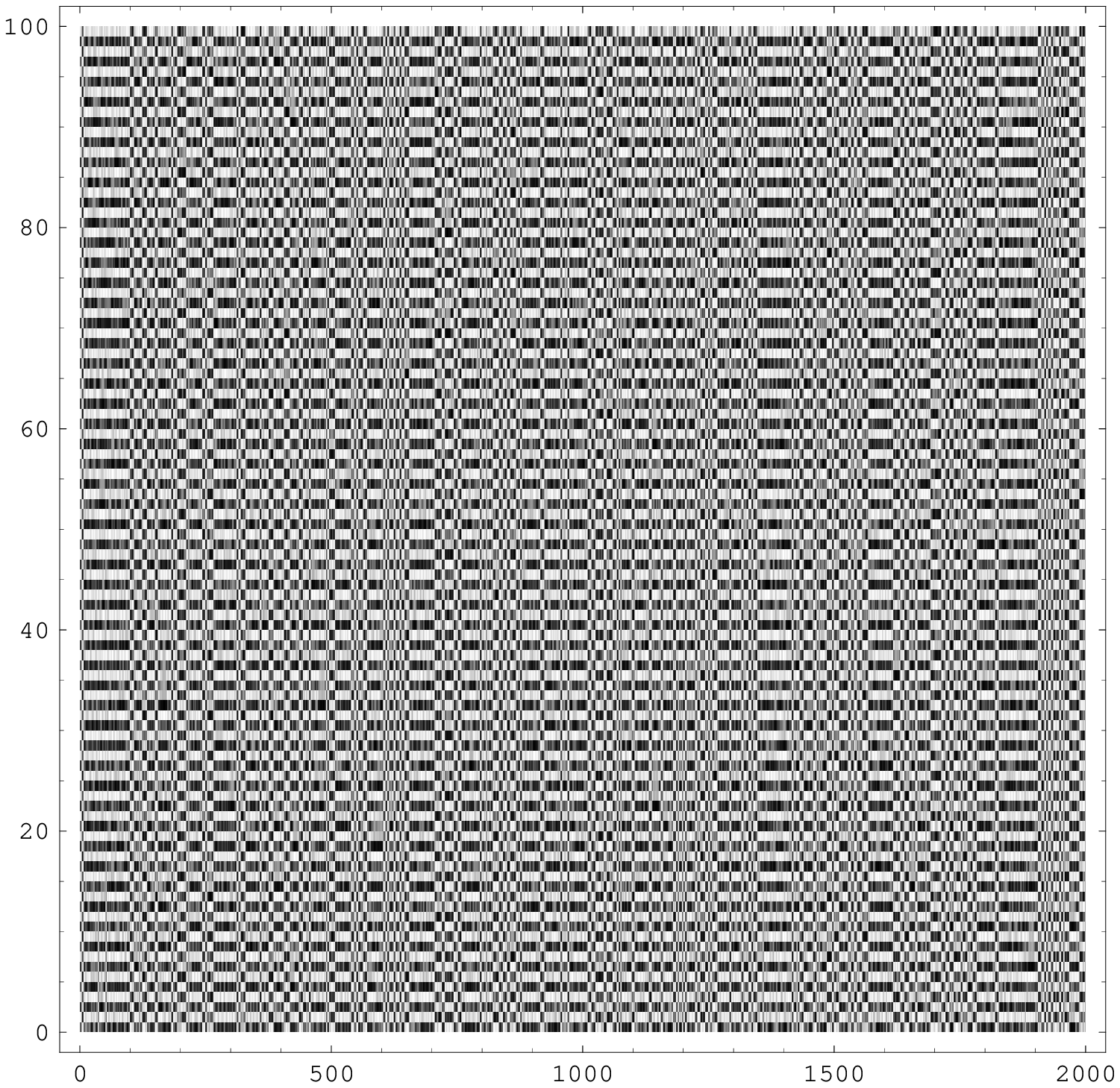}\\
(b) & \includegraphics[height=3.8in,width=4in]{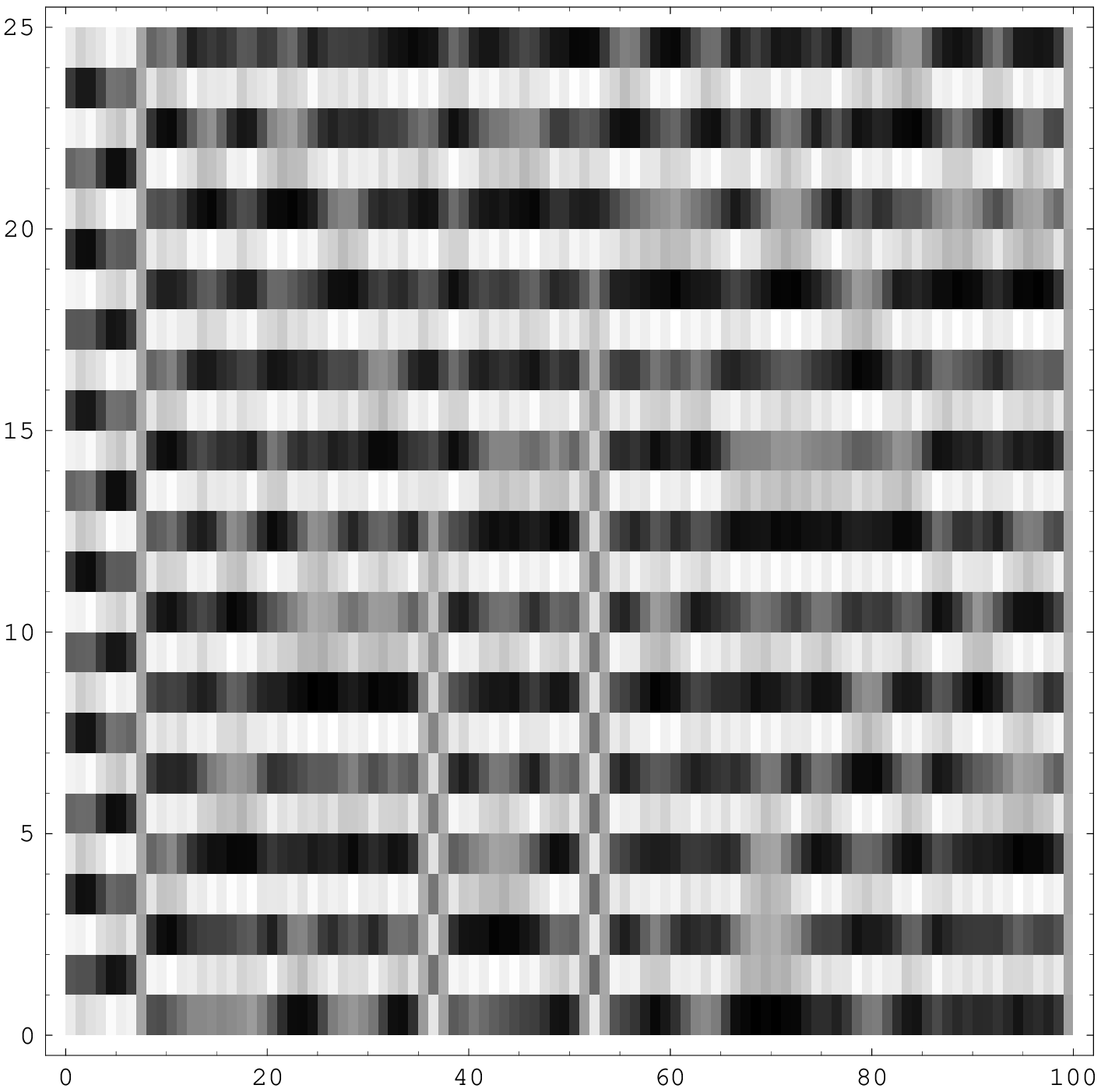}
\end{tabular}
\caption{Spatiotemporally intermittent behaviour seen in the inhomogenously coupled logistic map at $\gamma=0.66$ and $\epsilon=0.39$. (a) $L=2000$, last 100 iterations are shown after discarding $20,000$ transients. (b) $L=500$. The last 25 iterations are shown after discarding $5000$ transients.}
\end{figure}
\end{center}


\begin{thebibliography}{99}

\bibitem{Kaneko}
K. Kaneko, Prog. Theor. Phys. {\bf 58}, 112 (1987).

\bibitem{kaneko1}
{\it Theory and Applications of Coupled Map Lattices}, edited
by K. Kaneko (John Wiley, England, 1993) and references therein.

\bibitem{Bohr}
{\it Dynamical systems Approach to turbulence} by T. Bohr, M.H. Jensen,
G. Paladin and A. Vulpiani,
Cambridge University Press, (1998).

\bibitem{Domany}
E. Domany and W. Kinzel, Phys. Rev. Lett. {\bf 53}, 311 (1984).

\bibitem{GL}
H. Chat\'e, Nonlinearity {\bf 7},185 (1994).
\bibitem{St}
M. G. Zimmermann, R. Toral, O. Piro and M. San Miguel,Phys. Rev. Lett.,
{\bf 85}, 3612, 2000.



\bibitem{Kapral}R. Kapral,  Chap. 5, pg. 135, in {\it Theory and
applications ofcoupled map lattices} ed. by K. Kaneko, (John Wiley)
(1993).


\bibitem{RB}
F. Daviaud, M. Dubois and P. Berge, Europhys. Lett. {\bf 9}, 441(1989).
\bibitem{RB1} S. Ciliberto and P. Bigazzi, Phys.Rev.Lett.{\bf 60}, 286
(1988).
\bibitem{PC}
S. Bottin, F. Daviaud, O. Dauchot, and P. Manneville,  Europhysics
    Letters, {\bf 43},  171, (1998).
\bibitem{TD}
M.M. Degen, I. Mutabazi and C.D. Andereck Phys.Rev. E{\bf 53} 3495
(1996).
\bibitem{Colovas} G. Colovas and C. David Andereck, Phys.Rev.E {\bf
55(3)}
2736 (1997).
\bibitem{CT}
A. Goharzadeh and I. Mutabazi, Eur. Phys. J. B. {\bf 19},157 (2001).

\bibitem{printer} S. Michalland, M. Rabaud and Y. Couder, Europhys.Lett.
{\bf 22} 17 (1993).

\bibitem{Pomeau}
Y. Pomeau, Physica D{\bf 23}, 3 (1986).



\bibitem{Rolf} J. Rolf, T. Bohr and M.H. Jensen, {\it Phys.
Rev. E} {\bf
  57} (1998) R2503.


\bibitem{Chate}
H. Chat\'e and P. Manneville, Physica D {\bf 32}, 409 (1988).


\bibitem{Houlrik}
J.M. Houlrik, {\it et al}, {\it
    Phys. Rev. A} {\bf 41} 4210 (1990).

\bibitem{Grassberger}
P. Grassberger and T. Schreiber, {\it Physica D}
  {\bf 50}, 177 (1991).



\bibitem{Janaki}

T.M. Janaki, S. Sinha and  N. Gupte,
Phys. Rev. E {\bf 67}, 056218 (2003).


\bibitem{Faraday}
J. Gollub and S. Ramashankar, {\it New Perspectives in Turbulence} ed. 
S. Orszag and L. Sirovich, Springer, Berlin (1991).

\bibitem{electroconvection}
S. Nasuno and S. Kai, Europhys. Lett. {\bf 14}, 779 (1991).

\bibitem{Bohr1}
 T. Bohr, M. van Hecke, R. Mikkelson and M. Ipsen,  Phys. Rev. Lett., {\bf 86},
  5482 (2001).


\bibitem{Mikkelson}

R. Mikkelson, M. van Hecke and T. Bohr, arXiv: cond-mat 0207208.


\bibitem{zahera}

Z. Jabeen and N. Gupte, {\it Phase diagram of the sine circle map
lattice}, arxiv:nlin.CD/0502053.

\bibitem{FN}
Non-universal spreading exponents are seen only for the cases where the 
initial state is homogeneous with symmetrically placed seeds leading to
strictly symmetric spreading. However, very small departures from
homogeneity are sufficient to restore the DP exponents. 



\bibitem{FN2}
The sine circle map system also shows a reasonably smooth dependence of
the order parameter $m$ on $\epsilon$ when the order parameter is
approached from below $\epsilon_c$, unlike the case of the logistic map 
discussed in \cite{Grassberger}. This again contributes to the
appearance of DP-like behaviour in the system.





\bibitem{pinn}
R.O. Grigoriev, M.C. Cross and H.G. Schuster, Phys. Rev. Lett. {\bf 79},
2795(1997).



\bibitem{AS}

A. Sharma and  N. Gupte,
Phys. Rev. E {\bf 66}, 036210 (2002).

\bibitem{AS1}
A. Sharma and  N. Gupte,
International Journal of Bifurcations and Chaos, {\bf  12}, 1363
(2002).





\bibitem{Rupp}
P. Rupp, R. Richter and I.  Rehberg, arXiv:
  cond-mat/0201308.




\end{thebibliography}
\end{document}